\documentstyle[multicol,aps,epsf]{revtex} 
 
\tighten       

\begin{document}

\title{Asymptotic analysis of the Gunn effect with realistic
boundary conditions.}
\author{L. L. Bonilla}
\address{
Escuela Polit\'{e}cnica Superior\\
Universidad Carlos III de Madrid\\
Butarque 15, 28911 Legan\'{e}s, Spain}

\author{I. R. Cantalapiedra}
\address{Departament de F\'{\i}sica Aplicada\\
Universitat Polit\'{e}cnica de Catalunya\\
Gregorio Mara\~{n}\'{o}n 44, 08028 Barcelona, Spain}
\author{ G. Gomila and J. M. Rub\'{\i}}
\address{Department de F\'{\i}sica Fonamental \\ 
	Universitat de Barcelona\\
 Diagonal 647, 08028 Barcelona, Spain}

\date{\today}
\maketitle
\parskip 2ex

\begin{abstract}
A general asymptotic analysis of the Gunn effect in n-GaAs under
general boundary conditions for metal-semiconductor contacts is presented.
Depending on the parameter values in the boundary condition of the 
injecting contact, different types of waves mediate the Gunn effect.
The periodic current oscillation typical of the Gunn effect
may be caused by moving charge-monopole accumulation or
depletion layers, or by low or high-field charge-dipole solitary waves.
A new instability caused by multiple shedding of (low field) dipole
waves  is found. In all cases the shape of the current oscillation is
described in detail: we show the direct relationship between its major
features (maxima, minima, plateau's, \ldots) and several critical currents
(which depend on the values of the contact parameters).
Our results open the possibility of measuring contact parameters from
the analysis of the shape of the current oscillation. 
\end{abstract}

\draft{PACS: 05.45+b, 72.20.Ht, 85.30.Fg} 

\begin{multicols}{2}
\narrowtext

\section{Introduction}
The Gunn effect appears in many semiconductor samples presenting negative 
differential resistance and subject to voltage bias conditions 
\cite{Gunn,kro72,Grubin1,Grubin2,Kahn,Bergmann,Maan}. It 
consists of a periodic shedding of pulses of the electric field
at the injecting contact, which then progress and are annihilated at
the receiving contact. As a result there appears a periodic oscillation 
of the current through a passive external circuit attached to the
semiconductor. Under different conditions, the current self-sustained
oscillation may be caused by the motion of charge accumulation layers
(charge monopoles) \cite{kro72}, not by the usual electric field pulses 
which are charge dipoles. Most of the experiments on the Gunn effect in 
different materials take place in samples with attached planar 
contacts, so that the wave motion may be safely assumed to be
one-dimensional. Despite the vast literature on the Gunn effect, it
is surprising that many basic questions remain poorly understood.
For example, given a description of the charge transport in the bulk
semiconductor (say at the level of drift-diffusion and rate equations),
which are the proper boundary conditions for given contacts and how they 
affect the self-sustained current oscillations. The first question has 
been addressed in a companion paper, Ref.~\cite{companion}, while 
the second will be answered here. 

Until recently, when confronted with the Gunn effect, theorists
would resort to computer simulations of more or less complicated
models (which were supposed to reflect the physics of a given 
semiconductor), and would then explain qualitatively their numerics.
Resort to special solutions valid for infinite semiconductors at 
constant current bias conditions \cite{Grubin1}, or to extrapolations
of Kroemer's NL criterion \cite{photogunn} were often
used to interpret the simulation results. This left the processes of
generation and annihilation of domains at the contacts (and in fact
it also left out the dynamics of wavefronts and pulses) outside 
theoretical considerations.
Concerning asymptotic descriptions of the Gunn effect which delve deeper 
than just numerical simulations of drift-diffusion models, some progress
has been made recently \cite{HB,BKMV,BHHKV,univ}. These works propose 
asymptotic descriptions of the Gunn effect exploiting the
fact that this effect is seen most clearly in semiconductors
having a large value of the product of sample length times doping 
(basically a dimensionless length). The role of the ``N-L product'' 
in the analysis of the Gunn instability was already discovered by 
Kroemer~\cite{kro72} and exploited to study the linear stability 
(small signal analysis) of stationary solutions by many authors \cite{Grubin1}.
It was recognized only much later that in the limit of large dimensionless 
length (N-L product) it is possible to describe asymptotically both the 
onset \cite{onset} and the fully developed Gunn instability \cite{HB}.
In this asymptotic limit the processes of repeatedly generating a new wave 
(a charge monopole or dipole domain) at the injecting contact, the motion 
of the wave and its annihilation at the receiving contact may be well 
separated. Then they can be analyzed and combined to fully describe the 
Gunn effect. In particular, the effect of contacts on these
processes and in determining the shape of the current oscillation
can be clearly stated.
In this paper we use our asymptotic theory to study Kroemer's
model for n-GaAs under b.c.\ corresponding to ideal MS contacts. We find
that these b.c. give rise to a multivalued control current-field 
characteristic at the injecting contact. The asymptotic analysis shows that 
the Gunn effect can be mediated by both charge monopole or dipole domains
according to the values of the contact parameters. Shedding of new charge 
dipole waves from the injecting contact is adiabatic in clear distinction
with what happens if the control characteristic of the contact is
single-valued \cite{BHHKV,univ}. In the later case (analyzed for a p-Ge 
model in  Ref.\ \cite{BHHKV}) the charge dipole pulses are created very
rapidly at the injecting contact and they advance and grow simultaneously
(see also \cite{HB}), whereas for multivalued control characteristic
the boundary layer at the injecting contact grows adiabatically to a 
much greater size before a new pulse can be shed. These facts may
determine appreciably the shape of the current
oscillations. Our analysis could be extended to more complex 
models also displaying the Gunn effect, \cite{BHHKV,univ,buttiker}. 
Depending on the values of the parameters characterizing the injecting 
contact and of the dc voltage bias, we find Gunn oscillations
mediated by charge accumulation and depletion monopole wavefronts,
and high and low field charge dipole domains. We also find narrow
regions in the parameter space where multiple shedding of dipole
domains at the injecting contact occurs. We have thus found a
characterization of all possible dynamic behaviors which an
ideal metal-semiconductor contact would give rise to in the
Kroemer model for the Gunn effect. This opens the possibility
of extracting information about the contacts from the analysis of
the Gunn oscillations themselves, a subject of considerable interest
for applied researchers.

The rest of the paper is as follows. In Section II we present 
Kroemer's model and the boundary conditions for metal-semiconductor
contacts discussed in the companion paper \cite{companion}. 
In Section III we present our 
asymptotic analysis of Kroemer's model 
and find that different types of Gunn effect are possible according
to the values of the bias and of certain dimensionless
parameters appearing in the b.c., $i_0$ and $\alpha_{0}$: charge
monopoles (moving charge depletion and
accumulation layers), high and low field solitary waves (charge dipoles),
multiple (low field) charge dipoles,
are predicted and confirmed by numerical simulations. Section V contains
our conclusions whereas the Appendices are devoted to different technical 
matters related to the main text.

\section{Kroemer's model and b.c. for metal-semiconductor contacts}
The unipolar drift-diffusion model for the Gunn effect proposed 
by Kroemer \cite{kro72,Kroemer},
 is generally accepted to provide a
rather complete description of the main features of this effect.
In the dimensionless units described in Ref. \cite{companion},
 Kroemer's model is
\begin{eqnarray}
{\partial E\over\partial t} + v(E)\, \left( {\partial E\over\partial x} 
+ 1\right) - \delta {\partial^{2} E\over\partial x^{2}} = J,\label{eq}\\
{1\over L}\, \int_0^L E(x,t)\, dx = \phi. \label{bias}
\end{eqnarray}
Eq.~(\ref{eq}) is Amp\`ere's law which
says that the sum of displacement current and drift-diffusion current is
equal to the total current density $J(t)$. It can be obtained by 
differentiating the Poisson equation, $\partial E/\partial x = n-1$, with
respect to time, substituting the charge continuity equation
$\partial n/\partial t + \partial j(x,t)/\partial x = 0$ [the electron
current density is of the drift-diffusion type: $j(x,t) = n v(E) - 
\delta\, \partial n/\partial x$], and then integrating the result with 
respect to $x$. The electron velocity is assumed to be 
N-shaped. For specific numerical calculations we shall use
Kroemer's curve~\cite{Kroemer}
\begin{equation}
v(E) = E\, {1 + B E^{4}\over 1 + E^{4}},\label{v}
\end{equation}
(it has a maximum $v_M>0$ at $E_M>0$ followed by a minimum $0<v_m <
v_M$ at $E_m>E_M$) and the electron difusivity $\delta$ to be constant. 
The dc 
bias $\phi$ is the average electric field on the semiconductor sample. 
Eqs.~(\ref{eq})-(\ref{bias}) need to be solved with an appropriate initial 
field profile $E(x,0)$ and subject to the corresponding b.c.  
For ideal metal-semiconductor the following mixed boundary conditions
have been derived in \cite{companion}
\begin{eqnarray}
\frac{\partial E}{\partial x}(0,t) & = & \alpha_{0} \left(i_{0} - J(t) + 
\frac{\partial E}{\partial t}(0,t)\right)\,, \label{bc02} \\
\frac{\partial E}{\partial x}(L,t) & = & \alpha_{L}  \left(i_{L} + J(t) - 
\frac{\partial E}{\partial t}(L,t)\right)\, . \label{bcl2}
\end{eqnarray}
where $i_{i}$ and $\alpha_{i}$, ($i=0,L$), are dimensionless parameters which
are a combination of the semiconductor effective density of states, contact
barrier height, Richardson's constant, doping and 
temperature (see \cite{companion}).
In what follows, $i_{i}$ will be assumed to be positive because 
the physically interesting  phenomena (including the 
usual Gunn effect mediated by high field domains) are observed
for these values of $i_{i}$ (see phase diagram in \cite{companion}).

For typical n-GaAs data, $\delta\ll 1$ and $L\gg 1$ \cite{HB}. In this 
limit, we shall find approximate solutions to the initial boundary value
problem (\ref{eq})-(\ref{bcl2}) for $E(x,t)$ and $J(t)$. 
Strictly speaking, the simple asymptotic description that follows holds 
in the limit $L\to\infty$, even when $\delta =O(1)$~\cite{univ}. Assuming 
$\delta\ll 1$ just simplifies the description of the traveling waves of 
electric field in the semiconductor through the use of characteristic 
equations and shock waves~\cite{HB,Knight,SIAM91}. For example, it is
shown in Appendix \ref{carac} (by using the method of characteristics)
that the boundary condition (\ref{bc02}) implies that the electric field 
at the injecting contact, $E(0,t) = E_0(t)$, obeys the following equation:
\begin{eqnarray}
{dE_{0}\over dt} = J - j_c(E_0), \label{E_0}	 \\
j_{c}(E) = \frac{(1+\alpha_{0} i_{0}) v(E)}{1+\alpha_{0} v(E)}.\label{jc}
\end{eqnarray} 
These expressions constitute a Dirichlet boundary condition for the 
electric field which contains the same information as the mixed condition 
(\ref{bc02}). The contact curve, $j_{c}(E)$, presents two extrema, a minimum,
$j_{cm} = j_c(E_m)$, and a maximum, $j_{cM}= j_c(E_M)$, at the same 
field values as the electron velocity curve $v(E)$. $j_c(E)$ tends 
to $j^{sat}_{0}=\alpha_{0}^{-1}+i_{0}$ for high electric fields. As we shall 
see below, during most of an oscillation period, $j_c(E_0) \sim J$, and 
this expression yields a multivalued contact-characteristic curve relating 
the field at the injecting contact to the actual value of the current density.

To take advantage of the large-length limit, we will use the following 
rescaled time and length,
\begin{equation}
\epsilon = \frac{1}{L}\,,\quad s = {t\over L}\,,\quad
\quad y = {x\over L}\, .\label{slow}
\end{equation}
Then Eqs.~(\ref{eq})-(\ref{bias}) become 
\begin{eqnarray}
J - v(E) = \epsilon \left( {\partial E\over\partial s} + v(E)\, 
{\partial E\over\partial y} \right) - \delta\epsilon^2 {\partial^{2} 
E\over\partial y^{2}} \,,\label{s-eq}\\
\int_0^1 E(y,s)\, dy = \phi. \label{s-bias}
\end{eqnarray}
Notice that the boundary condition (\ref{E_0}) becomes
\begin{equation}
J - j_c(E) = \epsilon {dE_{0}\over ds}\,, \label{s-E_0}
\end{equation}
in the limit $\delta\ll 1$.

\section{Asymptotics of the Gunn effect}
\begin{figure}     
\centerline{
    \epsfxsize=8 cm
    \epsffile{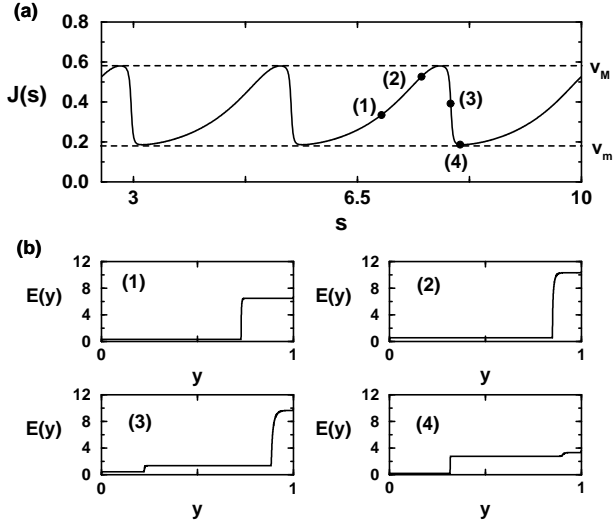}}
\caption{Gunn effect mediated by monopole wavefronts (moving accumulation
charge monopoles). (a) Dimensionless 
current density $J(s)$. Parameter values are $L=800$, $i_{0} = 1.35$, 
$\alpha_{0}=0.8$, 
and  $\phi=2$, for which $j_{cm}<v_{m}$ and $j^{sat}_{0}>v_{m}$,
where $j^{sat}_{0}$ corresponds to the saturation current 
$j_{0}^{sat}= \alpha^{-1} + i_{0}$. The maximum and minimum values
of the oscillation correspond to the maximum and minimum values
of the $v(E)$ curve, $v_{M}=0.58$ and $v_{m}=0.18$, respectively.
(b) The corresponding electric 
field profiles $E(y,s)$ evaluated at the times marked in Part 
(a) of this figure.}
\label{monacu}
\end{figure}

\begin{figure}   
\centerline{
    \epsfxsize=8 cm
    \epsffile{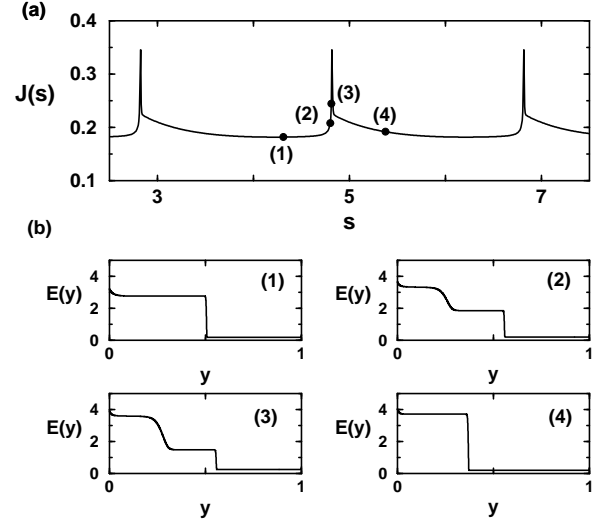}}
\caption{Gunn effect mediated by monopole wavefronts (moving depletion
charge monopoles). (a) Dimensionless 
current density $J(s)$. Parameter values are $L=800$, $i_{0} = 0.135$,
 $\alpha_{0}=0.8$, 
and  $\phi=1.5$, for which $j_{cM}>v_{M}$. (b) The corresponding electric 
field profiles $E(y,s)$ evaluated at the times marked in Part 
(a) of this figure. }
\label{mondep}
\end{figure}

In a previous paper \cite{companion} we have analyzed the 
stationary solutions of Kroemer's model with metal-semiconductor contacts 
and discussed their stability. No stable stationary solution is expected
for certain ranges of bias and the cathode contact parameters $i_{0}$
and $\alpha_{0}$. In these circumstances, the Gunn effect 
mediated by either moving charge monopoles or dipoles might appear. A 
rich phenomenology of propagating waves and current oscillations has been 
numerically observed for these parameter ranges. Among them, we have observed 
both high (Figs.\ \ref{diphig1}, \ref{diphig2}) and low (Fig.\ \ref{diplow}) 
field solitary waves (moving charge dipoles), multiple low field 
dipoles (Fig.\ \ref{multi}), moving charge accumulation (Fig.\ \ref{monacu}) 
and charge depletion monopoles (Fig.\ \ref{mondep}). In 
Ref.\\cite{companion}, we have identified the critical currents determining 
which type of waves mediate the current oscillation (in the limit $\delta\ll 
1$). They are related to the boundary conditions
in the following way: 
\begin{itemize}
\item If $j_{cM}>v_{M}$, the Gunn effect is mediated by moving charge 
accumulation monopoles [$j_{cM}$ and $v_M$ are the local maxima of the contact
current and electron velocity curves, both reached at $E=E_M$].
\item  If  $j_{cm}<v_{m}$, and $j^{sat}_{0}> v_m$, the Gunn effect is 
mediated by moving charge depletion monopoles [$j_{cm}$ and $v_m$ are the 
local minima of the contact current and electron velocity curves, both 
reached at $E=E_m$; $j^{sat}_{0}= \alpha_0^{-1} + i_0$ is the value at which 
$j_c(E)$ saturates at high electric fields].
\item If $v_{m}<j_{cm}<j_{cM}<v_{M}$, moving charge dipoles
mediate the Gunn effect (see Ref.\cite{companion} for more details). 
\end{itemize}

As shown in the figures, there are several stages in 
each period of the current oscillation corresponding to the processes
of generation, propagation and annihilation of electric field domains.
Each stage has its own time and space scales, 
and hence the oscillation can be suitably described by means of a
matched asymptotic analysis. For instance, the annihilation of wavefronts 
takes place on a fast time scale compared to that governing 
wavefront propagation. Thus on the time scale of wavefront propagation,
the annihilation of wavefronts is a quasi-instantaneous process during
which the time derivative of the current density changes appreciably
while the current itself, $J(s)$, does not change. On the other hand, 
the generation of fronts takes place adiabatically on a much slower 
time scale comparable to wavefront propagation. In this stage both 
$J(s)$ and its derivative change appreciably. This is quite different
from the fast generation of fronts and pulses observed for other 
types of boundary conditions \cite{BHHKV}. 

As long as these different processes take place on different
time and space scales, there are different stages of the oscillation
which can be analyzed separately. This happens for certain bias ranges.
We shall then construct the approximate electric field and current density
solutions by means of matched asymptotic expansions. A detailed description
of a period of the current oscillation will be then obtained. For other
bias values, several processes occurs almost simultaneously (e.g., 
the annihilation of a wavefront may occur during the process of
detachment of another wavefront from the injecting boundary layer 
for low bias values). This complicates the asymptotic description 
without adding much to our physical understanding, so that we 
will omit the details.

Note that in the limit $\epsilon =1/L \ll 1$, the solutions of 
Eqs.~(\ref{s-eq})-(\ref{s-bias}) are piecewise constant: on most of the 
$y$-interval, $E$ is equal to one or another of the 
zeros of $v(E)-J$ [notice that this equation may have three
zeroes, we denote by $E_{1}<E_{2}<E_{3}$], separated by transition 
layers that connect them. At 
$y=0$ and $y=1$ there are boundary layers (quasi-stationary most of the 
time), which we call {\em injecting and receiving layers}, respectively.
It can be seen that the  propagation of fronts turns out to be a 
quasi-stationary process while the generation is not. Thus, two different
asymptotic approaches will be used: one for the description of the
quasistationary propagation of fronts and the other for their 
(non-quasistationary) generation.

\subsection{Quasistationary propagation of wavefronts}
In this section we will present the asymptotics of the quasistationary
propagation of wavefronts, and of the corresponding time evolution
of the density current. 

\begin{figure}   
\centerline{
    \epsfxsize=8 cm
    \epsffile{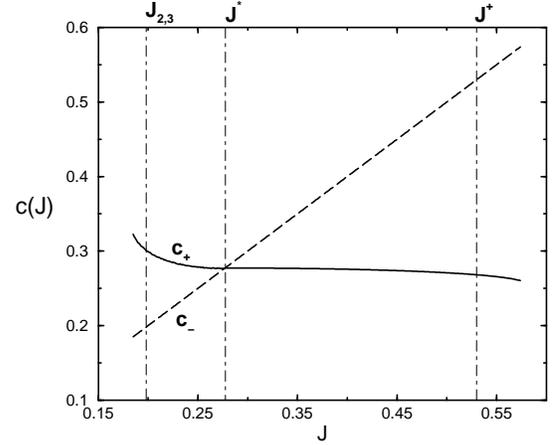}}
\caption{Velocities $c_+$ and $c_-$ of the heteroclinic wavefronts as functions 
of the current density. Notice that the lines intersect at $J^*=0.28$. We have 
also marked the currents $J^{\dag}=J_{2,1} =0.53 $ at which $2c_+ = c_-$ and 
 $J_{2,3} =0.20 $ at which $2c_+ = 3c_-$.}
\label{c-pm}
\end{figure}

Wavefronts are moving transition layers connecting regions of the sample
where the electric field is spatially uniform. In order to describe their
quasistationary propagation we will proceed as follows. 
We assume that $E(y,s)$ is either $E_1(J)$ or $E_3(J)$ outside 
boundary layers and wavefronts.  Let the wavefront located 
at $y=Y(s)$ move with velocity $c = dY/ds$. For each value of $J$, the
wavefront advances with speed $c=c_+(J)$, if it connects $E=E_1(J)$ [$y<Y(s)$] 
to $E_3(J)$ [$y>Y(s)$], or with speed
$c=c_-(J)$, if it connects $E=E_3(J)$ [$y<Y(s)$] to $E_1(J)$ [$y>Y(s)$].  
To find the inner structure of a wavefront and its speed, we introduce
a coordinate $\xi = \epsilon^{-1}[y-Y_i(s)]$ moving with the wavefront.
Then we need to solve the following problem for the equation 
\begin{equation}
\frac{dE}{d\xi} = F;\quad\quad 	\frac{dF}{d\xi} = {[v(E) - c]\, F + v(E) - J
\over\delta}\,,			\label{eq:phas}
\end{equation}
obtained from Eq.\ (\ref{s-eq}):
{\em  Find the unique value 
$c=c_+(J)$ [resp. $c_-(J)$] such that there is a solution of (\ref{eq:phas}) 
with $E(-\infty) = E_1(J)$ and $E(\infty) = E_3(J)$ [resp,\ $E(-\infty) = 
E_3(J)$ and $E(\infty) = E_1(J)$]}. The solution of this problem will
provide both the speed and inner structure of the wave front.
In terms of the phase plane (\ref{eq:phas}), the previous problem is
equivalent to find $c=c_+(J)$ 
so that there is a heteroclinic orbit connecting the saddle point $(E_1,0)$ 
to the saddle point $(E_3,0)$ with $F>0$. Similarly $c=c_-(J)$ corresponds 
to a heteroclinic orbit connecting $(E_3,0)$ to $(E_1,0)$ with $F<0$. The 
functions $c_{\pm}(J)$ for our model
are depicted in Fig.\ \ref{c-pm}. Note that they intersect 
when $J=J^*$, given by
\begin{equation}
J^{*} = \frac{1}{E_{3}- E_{1}}\,\int_{E_{1}}^{E_{3}} v(E) \ dE,	
\quad c_{\pm}=J^*,		\label{J*}
\end{equation}
the so-called equal-areas rule~\cite{Grubin2}. 

In the limit $\delta\ll 1$, we can obtain explicit expressions for 
$c_{\pm}(J)$ and for the corresponding wavefronts as we now 
recall~\cite{HB,SIAM91}. The wavefront (moving depletion layer) joining 
$(E_3(J),0)$ and  $(E_1(J),0)$ may be approximated by the exact solution
of (\ref{eq:phas})
\begin{eqnarray}
 E(\xi) = - \xi,\quad\quad F(\xi) = -1,\quad\quad c_-(J) = J,	\label{A1}
\end{eqnarray}
which holds for any value of $\delta$ plus two corner layers of width
$O(\sqrt{\delta})$ (on the $\xi$ scale) at $\xi = \pm (E_3- E_1)/2$. See 
Ref.\ \cite{SIAM91} for an explicit calculation. The width of this 
wavefront on the $\xi$ length scale is $(E_3-E_1) + O(\sqrt{\delta})$, 
which yields $y-Y_i(s) = O(\epsilon)$ on the large length scale. The velocity 
of this wavefront is $c_-(J) = J$. 

The other wavefronts (moving accumulation layers) can be constructed by
matched asymptotic expansions in the limit $\delta\ll 1$ and their 
velocities $c_+(J)$ and shapes depend on whether $J$ is larger or
smaller than $J^*$. These wavefronts are composed of a shock wave joining
two field values $E_-$ and $E_+$ (at least one of them should be equal to
$E_i$, $i=1,3$) plus a tail region which moves rigidly with the same 
velocity as the shock \cite{HB}. The inner structure of the shock wave
(for very small but not zero $\delta$) can be a quite complicated 
triple-deck set of boundary layers \cite{SIAM91}. Let $V(E_+,E_-)$ be the 
velocity of the shock wave given by the equal-areas
 rule~\cite{HB,Knight,SIAM91}
\begin{equation}
V(E_+,E_-) = 	\frac{1}{E_{+}- E_{-}}\,\int_{E_{-}}^{E_{+}} v(E) \ dE.	
		\label{A2}
\end{equation}
We now have~\cite{HB}:
\begin{enumerate}
\item If $J\in (v_m,J^*)$, $E_+ = E_3(J)$ whereas $E_-$ is calculated as
a function of $J$ by imposing the condition that the tail region to the
left of the shock wave moves rigidly with it 
\begin{equation}
V(E_3,E_-) = 	v(E_-).			\label{A3}
\end{equation}
Solving simultaneously (\ref{A2}) and (\ref{A3}) (with $E_+ = E_3$), we find 
both $E_-$ and $c_+ = v(E_-)>J$ as functions of $J$. To the left of the shock
wave (in the tail region) the field satisfies the (approximate) boundary 
value problem 
\begin{eqnarray}
[v(E) - c_+(J)]\, {dE\over d\xi} = J - v(E),\quad \quad \xi < 0, \nonumber\\
E(-\infty) = E_1(J),\quad\quad  E(0)=E_-(J).			\label{A4}
\end{eqnarray}
\item If $J\in (J^*,v_M)$, $E_- = E_1(J)$ whereas $E_+$ is calculated as
a function of $J$ by imposing the condition that the tail region to the
right of the shock wave moves rigidly with it 
\begin{equation}
V(E_+,E_1) = 	v(E_+).			\label{A5}
\end{equation}
Solving simultaneously (\ref{A2}) and (\ref{A5}) (with $E_- = E_1$), we find 
both $E_+$ and $c_+ = v(E_+)<J$ as functions of $J$. To the right of the shock
wave (in the tail region) the field satisfies the (approximate) boundary 
value problem 
\begin{eqnarray}
[v(E) - c_+(J)]\, {dE\over d\xi} = J - v(E),\quad \quad \xi > 0, \nonumber\\
E(0)=E_+(J),\quad\quad E(\infty) = E_3(J).			\label{A6}
\end{eqnarray}
\end{enumerate}
At $J=J^{\ast}$ we have $E_-=E_1$, $E_+=E_3$, and $c_+=J^*$. It is not hard
to prove that $c_+(J)$ is a decreasing function. The functions $c_{\pm}(J)$
are depicted in Fig.\ \ref{c-pm}. 

Notice that the inner structure of the shock wave has width $O(\sqrt{\delta})$
on the $\xi$ scale while the total width of the wavefront is $O(1)$ on the 
$\xi$ scale and $O(\epsilon)$ on the $y$ scale. In the limit $\delta\ll 1$, 
the structure of the wavefronts is thus one-sided: the wavefront is a 
discontinuity preceded or followed by a tail region. 

In conclusion, the propagation of a single wavefront can be described by giving
the position, $Y(s)$, and velocity, $c_{\pm}(J(s))$,
of the front and the values of the electric field on its left and right
hand sides, $E_{i}(J(s))$. When more than a single wavefront are
co-moving inside the sample, the same description applies to each of them.
All of the magnitudes involved in this description depend on time $s$ 
only through the instantaneous value of the
current $J(s)$. This fact, together with the fact that the voltage must 
remain constant all the time, can be used to derive a simple closed equation
describing the time evolution of the current density
during the propagation stages.  
As an example of this result, 
let us consider the case a single propagating dipole (two fronts) 
(see Fig.\ \ref{diphig1}b.1). In this case, neglecting
 transition and boundary layers, we have:
 $E(y,s) = E_{1}(J(s))$ for $0<y<Y_{1}(s)$ and $Y_{2}(s)<y<1$ and
 $E(y,s) = E_{3}(J(s))$ for 
 $Y_{1}>y>Y_{2}(s)$. For the voltage we have:
 \begin{eqnarray}
 \phi = E_1(J(s)) &+& [E_3(J(s)) - \nonumber \\
                  & & E_1(J(s))]\, (Y_2(s) - Y_1(s)) + O(\epsilon). \label{d1}
\end{eqnarray}
By using the fact that the voltage is fixed, 
 we can obtain an equation for $J(s)$ by differentiating the bias 
condition (\ref{d1}) with respect to $s$. By noting that 
\begin{eqnarray}
v(E_i) = J \Longrightarrow {dE_{i}\over dJ} = 	\frac{1}{v'(E_{i}(J))}\,,
\nonumber\\	
\quad {dY_{1}\over ds} = c_{+}(J), \quad {dY_{2}\over ds} = c_{-}(J), \label{d2}
\end{eqnarray}
the following simple closed equation for the current is obtained 
\begin{eqnarray}
{dJ\over ds} = 	A(J)\, [c_+ (J) - c_- (J)]\,,		\label{d3}\\
A(J) = \frac{(E_{3}-E_{1})^{2}}{{\phi-E_{1}\over v'_{3}} +{E_{3}-\phi
\over v'_{1}}} > 0,\label{d4} 
\end{eqnarray}
where $v'_i \equiv v'(E_i(J))$ ($i=1,3$). 
This equation for $J$ holds as long as the electric field profile
consist of a single propagating dipole. A simple analysis 
of Eq.(\ref{d3}) demonstrates that $J$ tends to $J^\ast$ 
(for which $c_+ (J)=c_- (J)$) 
exponentially fast, starting from a certain value of the current $J(0)$.
This result 
explains, in particular, why the solitary wave responsible for the Gunn 
effect moves at approximately constant velocity $c_\pm = J^\ast$ and constant 
current far from the contacts. 

A similar procedure can be applied to derive the corresponding closed equation
for the current when different types of propagating domains are present. 
In general we will have $n_{+}$ fronts moving at velocity
$c_{+}$ and $n_{-}$ moving at $c_{-}$, with $|n_{+}-n_{-}| = 0,1$. 
For these
cases the current evolves following the equation
\begin{equation}
{dJ\over ds} = 	A(J)\, [n_{+} c_+ (J)- n_{-} c_- (J)]\,. \label{dng}
\end{equation}
As before, this equation holds as long as the propagating profile 
consists of
$n_{+}$ fronts propagating at $c_+$ and $n_{-}$ at $c_{-}$.
Two typical cases can be considered in analyzing Eq.\ (\ref{dng}): (i)
either $n_{+}$ or $n_{-}$ are equal to zero and (ii) $n_{\pm}\neq 0$, 
both numbers are different from zero. In the first case, we have either
\begin{equation}
{dJ\over ds} = 	A(J)\, c_+ (J) \,, \hspace{1 cm}  n_{+}=1 \, , n_{-}=0
\label{dn10}
\end{equation}
or
\begin{equation}
{dJ\over ds} = - A(J)\, c_- (J) \,, \hspace{1 cm}  n_{+}=0 \, , n_{-}=1
\label{dn01}
\end{equation}
and the current will either increase, Eq.\ (\ref{dn10}), or decrease,
Eq.\ (\ref{dn01}), with time. In the second situation, the current follows 
the general relation Eq.\ (\ref{dng}) with the corresponding values of 
$n_{\pm}$: the current evolves towards the fixed point $J=J_{n_{+},n_{-}}$, 
satisfying $n_{+} c_{+}(J_{n_{+},n_{-}}) - n_{-} c_{-}(J_{n_{+},n_{-}})=0$, 
when such a point exists. A particular example of this behavior has been 
explained above for $J_{1,1}=J^{\ast}$.

These results describe the quasistationary propagation of fronts and the 
corresponding time evolution of the current density during these stages.
We will use them to interpret the results of the numerical simulations. 

\subsection{Generation of fronts}
As mentioned before, for our b.c.\ an appreciable part of a period of the 
current oscillation may be spent generating new fronts non-adiabatically.
Fronts of dipole domains are generated at the cathode whereas
monopole wavefronts appear somewhere in the middle of the sample. In what 
follows, we will focus on the description of dipole domains. Monopole 
wavefronts have been recently described in detail elsewhere\cite{BKMV}. 

A simple rule concerning generation of dipole domains can be formulated: 
when the current density, $J(s)$, crosses the maximum [resp.\ minimum] of
the contact curve, $j_{cM}$ [resp.\ $j_{cm}$], a front moving with
speed $c_{-}$ [resp.\ $c_{+}$], starts being formed {\em when its 
generation is compatible with the field value at the bulk after 
the injecting contact}. 

Let us now show why the previous rule holds. Suppose that the current 
reaches adiabatically in the slow time scale $s$ one of the critical values 
mentioned above, let us say $j_{cM}$. Then the field at $x=0$, 
given by Eq.\ (\ref{E_0}), can no longer be quasistationary,
the injecting layer becomes unstable and it 
starts shedding a new solitary wave. Let now $s_1$ be the earliest time 
at which $J=j_{cM}$, and the boundary field $E_0 = E_M$. After this time,
the disturbances $J-j_{cM}=O(\epsilon)$ cause $E_0$ to evolve to the third
branch of $j_{c}(E)$ on a time of order $s-s_1=O(\epsilon)$. The field in
the injecting layer, in turn, increases until a moving wavefront moving
with speed $c_-$ is formed. To describe this process, we adapt ideas 
developed for the analysis of the trap-dominated Gunn effect in 
p-Ge~\cite{BHHKV} to the present situation. An important difference 
is that the sharp increase in the contact field $E_0$ helps creating
the new wavefront but a new pulse is not shed in the fast time scale
$\sigma = (s-s_1)/\epsilon$: after the wavefront is created, the contact 
field varies on the third branch of $j_{c}(E)$ and the current has to 
decrease slowly [according to (\ref{dng}), on the $s$ time scale] until 
$E_0$ can jump back to values on the first branch of $j_{c}(E)$ when $J=
j_{cm}$.
 
To leading order, the field in the injecting layer solves the 
following semi-infinite problem whose derivation can be found in
Appendix \ref{shedding}: 
\begin{eqnarray}
{\partial E\over\partial\sigma} + v(E)\, \left( {\partial E\over\partial x} 
+ 1\right)  = J,
\label{eq-shed}\\
x>0,\quad\quad\quad -\infty < \sigma < +\infty\,,\nonumber\\
E(0,\sigma) = E_0(\sigma), \label{bc-shed}	
\end{eqnarray}
where $E_0$ solves (\ref{E_0}) with $J=j_{cM} + \epsilon J^{(1)}(\sigma)$,
and $J^{(1)}(\sigma)$ is given by 
\begin{eqnarray}
\left( {\partial \over\partial\sigma} + \beta \right) [J^{(1)} - h'(\sigma) 
- \alpha h] = - \gamma h, \label{eq:J1}
\end{eqnarray}
where $h(\sigma)$ is
\begin{eqnarray}
h(\sigma) = (E_3 - E_1)\, c_+ \, \sigma - \int_0^\infty [E(x,\sigma) 
- E_1]\, dx \nonumber\\
- \int_{-\infty}^0 [E(\xi) - E_1(j_{cM})]\, d\xi \nonumber\\
- \int_0^\infty [E(\xi) - E_3(j_{cM})]\, d\xi, \label{def:h}
\end{eqnarray}
 and formulas for the positive constants $\alpha$, $\beta$ and $\gamma$ 
may be found below, in Eqs.~(\ref{alpha}). The 
function $h(\sigma)$ is the area lost due to the motion of the old front 
during the time $\sigma$ minus the instantaneous excess area under the 
injecting layer minus the constant excess area under the old wavefront at
$Y=Y_1(s_1)$ [that is, under the heteroclinic orbit connecting 
$(E_1(j_{cM}),0)$ and $(E_3(j_{cM}),0)$].

As $\sigma\to -\infty$, we have to impose the following matching condition 
on an appropriate overlap domain:
\begin{eqnarray}
E(x,\sigma) - E_{stat}(x;J(s))\ll 1,\label{matching}\\
\mbox{as}\quad\quad \sigma\to - \infty,\quad\quad s \to s_1 -. \nonumber
\end{eqnarray}
Here $E_{stat}(x;J(s))$ is the quasistationary injecting layer solution 
of (\ref{eq:phas}) with $c=0$ such that $E_{stat}(0;J(s))$ satisfies 
(\ref{bc02}) and $E_{stat}(\infty;J(s)) = E_1(J(s))$ for $s<s_1$, $J(s_1) 
= j_{cM}$. Notice that the term $\epsilon J^{(1)}(\sigma)$ is needed so 
as to avoid that the solution of this problem stay indefinitely in the 
quasistationary field $E_{stat}(x;J(s))$. 

The solution of the previous semi-infinite problem reveals the 
growth of the field at the contact and inside the injecting layer until:
(i) $E_0$ becomes $E_{c3}(j_{cM})$ [$E_{c1}(J) < E_{c2}(J) < E_{c3}(J)$
are the three possible solutions of $J=j_c(E)$], (ii) $E(x,\sigma)$ 
increases to $E_3(j_{cM})$ as $x$ increases from $x=0$ and then (iii) it 
has the structure 
of a wavefront connecting $E_3(j_{cM})$ to $E_1(j_{cM})$. This wavefront
advances with velocity $c_-(j_{cM})$. 
The formation of a front when $J$ crosses $j_{cm}$ can be explained similarly.

\subsection{Putting the pieces together} 
In the two previous subsections, we have presented the basic features of the 
asymptotic description of the Gunn effect, namely, quasistationary front
propagation and the generation of new fronts. Now, we are in a position 
to put all the pieces together,
and describe a full period of current oscillation. We shall distinguish
three cases: high-field dipoles, low-field dipoles and monopoles.
\subsubsection{Dynamics of high-filed dipoles}
High-field dipole domains have been observed to appear when
$v_{m}<j_{cm}$ and $J^{\ast}<j_{cM}<v_{M}$. Then 
depending on the
value of $j_{cm}$ with respect to $J^{\ast}$, and of $j_{cM}$ with respect
to $J^{\dag} \equiv J_{2,1}$, different situations may occur.

\begin{figure}     
\centerline{
    \epsfxsize=8 cm
    \epsffile{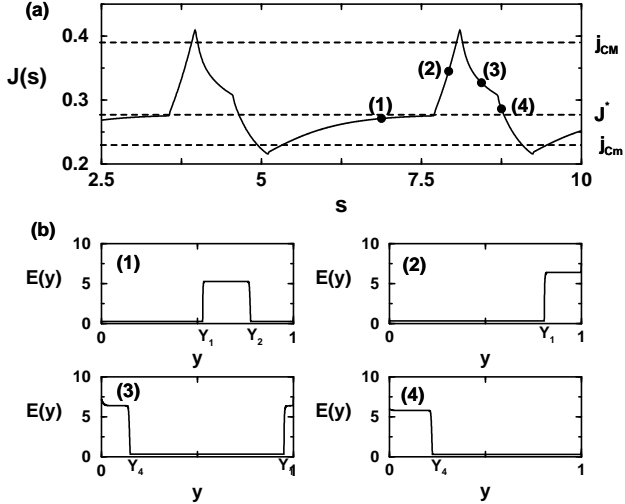}}
\caption{Gunn effect mediated by dipole solitary waves. (a) Dimensionless 
current density $J(s)$. Parameter values are $L=800$, $i_{0} = 0.3$,
 $\alpha_{0}=3.6$
and $\phi=1.5$, for which $v_{m}<j_{cm}<J^{\ast}<j_{cM}<v_{M}$. 
The minimum and maximum values of $J(s)$
 corresponds to $j_{cm}=0.23$ and $j_{cM}=0.31$, respectively whereas 
the plateau at intermediate values of $J(s)$ corresponds to the solution of 
$c_+(J)=c_-(J)$, $J^{\ast} =0.28$. 
(b) The corresponding electric 
field profiles $E(y,s)$ evaluated at the times marked in Part 
(a) of this figure. 
}
\label{diphig1}
\end{figure}

Let start considering the case
$v_{m}<j_{cm}<J^{\ast}<j_{cM}<J^{\dag}<v_{M}$. 
This situation corresponds to the propagation of high-field dipole 
domains as shown in Fig.\ \ref{diphig1}. We will assume the initial electric 
field configuration to correspond to a single propagating high field domain 
(Fig.\ \ref{diphig1}b.1). This configuration
corresponds to $n_{+}=1$ and $n_{-}=1$, and hence the current satisfies
Eq.\ (\ref{d3}), evolving from an initial value $J(0) \in (j_{cm}, J^{\ast})$,
towards the fixed point $J^{\ast} = J_{1,1}$. 
After a certain time, the wavefront located at $Y_2$ reaches the end
of the sample and disappears, producing an abrupt change in the time
derivative of the current.  
We have a new stage with $n_{+}=1$ and $n_{-}=0$, Fig.\ \ref{diphig1}b.2, 
governed by Eq.\ (\ref{dn10}).
Its solution increases until it surpasses the value $j_{cM}$.
At this point the injecting layer becomes unstable and it 
starts shedding a new front. The formation dynamics has been
explained in detail in the previous section. Then a new slowly varying 
stage begins. There are two leftover wavefronts, the old one located at 
$y=Y_1(s)$ [which advances toward $y=1$ with speed $c_+(J)$], and a new 
one located at $y=Y_4(s)$, moving with speed $c_-(J)$, and leaving behind 
a quasistationary field $E_3(J)$, see Fig.\ \ref{diphig1}b.3. Again, the field
configuration corresponds to $n_{+}=n_{-}=1$, with $J(s)$ following 
(\ref{d3}), and decreasing exponentially fast towards $J^*$. Before
$J$ reaches $J^*$, the front located at $Y_{1}$ reaches the end
of the sample and disappears, thereby producing a new abrupt change in the
time derivative of the current density. Then only the recently formed 
front [located at $Y_4(s)$] is present on the sample (Fig.\ \ref{diphig1}b.4), 
which corresponds to $n_{+}=0$ and $n_{-}=1$. $J(s)$ decreases according to 
Eq.\ (\ref{dn01}) until the minimum of the contact curve, $J = j_{cm}$,
is reached. After that, a front moving with speed $c_+$ is 
generated. The charge dipole wave thus created evolves adiabatically, and
the current density is described again by Eq.~(\ref{d3}), Fig.\ref{diphig1}b.1. 
We have come back to the initial situation, and one period of the current 
oscillation has been completed. 

It is worth noting that by means of this analysis some of the 
most relevant features
of the current oscillations, as the maximum and minimum currents, $J_{max}$ and
$J_{min}$,
have been identified with quantities related to the contact parameters,
$J_{max} \approx j_{cM}$ and $J_{min} \approx j_{cm}$, opening then
the possibility of determining the values of contact parameters
from the analysis of the current oscillations.

\begin{figure}     
\centerline{
    \epsfxsize=8 cm
    \epsffile{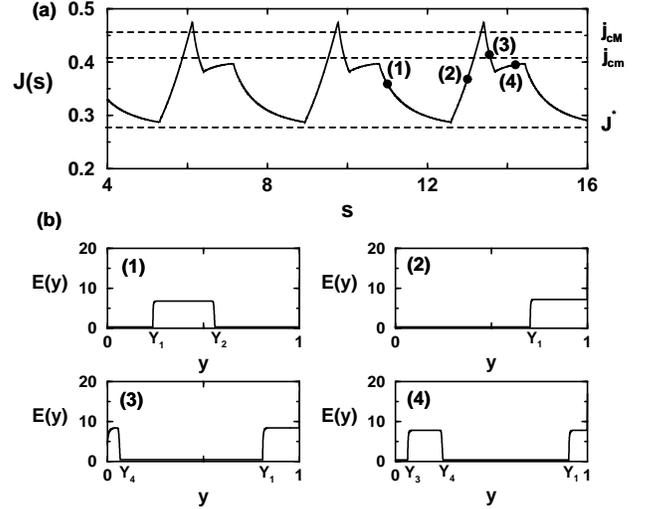}}
\caption{Gunn effect mediated by dipole solitary waves (a) Dimensionless 
current density $J(s)$. Parameter values are $L=1000$, $i_{0} = 0.45$,
 $\alpha_{0}=30$
and $\phi=2.4$, for which $v_{m}<J^{\ast}<j_{cm}<j_{cM}<v_{M}$.  
The maximum value of $J(s)$
 corresponds to $j_{cM}=0.46$, whereas 
the plateau at bottom of $J(s)$ corresponds to the solution of 
$c_+(J)=c_-(J)$, $J^{\ast} =0.28$. The value of $j_{cm}$ is $0.41$. 
(b) The corresponding electric 
field profiles $E(y,s)$ evaluated at the times marked in Part 
(a) of this figure. 
}
\label{diphig2}
\end{figure}

Other situations involving dipole domains have been
numerically identified. They are described by the same type of analysis. 
In what follows only the more relevant features
of these situations will be considered. Let us assume now
$v_{m}<J^*<j_{cm}<j_{cM}<J^{\dag}<v_{M}$.
This case corresponds again to high-field solitary waves (dipole domains, 
see Fig.\ \ref{diphig2}), but the current oscillations have a different shape.
The three first stages of the oscillation, Fig.\ \ref{diphig2}b.1,2,3,  
correspond to the propagation of a single dipole, annihilation of a front and
generation of new one. They are similar to the stages described above.
Now, however after the new front has been formed, Fig.\ \ref{diphig2}b.3, 
and the current is 
decreasing towards $J^*$, $J$ reaches the critical current, $j_{cm}$
(because $J^*<j_{cm}$). A new front moving with speed $c_{+}$ is then
created. After the formation process has finished, we have a configuration 
with two fronts moving at $c_{+}$ and one at $c_{-}$, that is, with $n_{+}=2$ 
and $n_{-}=1$ (Fig.\ \ref{diphig2}b.4). Hence the current satisfies [see Eq.\ 
(\ref{dng})],
\begin{eqnarray}
{dJ\over ds} = A(J)\, [2 c_+(J) - c_-(J)], \label{d7}
\end{eqnarray}
and it starts increasing again, trying to reach $J_{1,2} \equiv J^{\dag}$. Before this value 
may be attained, the old front located at $Y_{1}$, arrives at the receiving 
contact and disappears. We obtain again Eqs.\ 
(\ref{d3})-(\ref{d4}) and recover the initial situation. Thus a full period 
of the Gunn oscillation is again completed.

There are other possible situations for propagating 
high field domains, but we have not found them in our numerical simulations 
with the curve $v(E)$ considered in the present work. For example 
in the second case we have described above,
$v_{m}<J^*<j_{cm}<j_{cM}<J^{\dag}<v_{M}$, after the 
formation of the new front, Fig.\ \ref{diphig2}b.4, the current density 
could have reached the value $J^{\dag}$ {\em before} the old front 
located at $Y_{1}$ had arrived at the receiving contact. In such situation, 
a new front moving at $c_{-}$ could be formed, giving rise
to an electric field configuration with $n_{+}=n_{-}=2$. In this 
configuration the current would decrease again to 
$J_{2,2}=J_{1,1}=J^*$. We would then have a more complicated situation 
with two pulses and the dying wavefront simultaneously present in
the sample. Proceeding in a similar way, we could therefore observe 
simultaneous coexistence of several pulses for appropriate ranges of 
parameters. This situation (multiple shedding of high field domains)
has been numerically observed in other models \cite{BHHKV,univ}.

\subsubsection{ Dynamics of low-field dipoles}
\begin{figure}    
\centerline{
    \epsfxsize=8 cm
    \epsffile{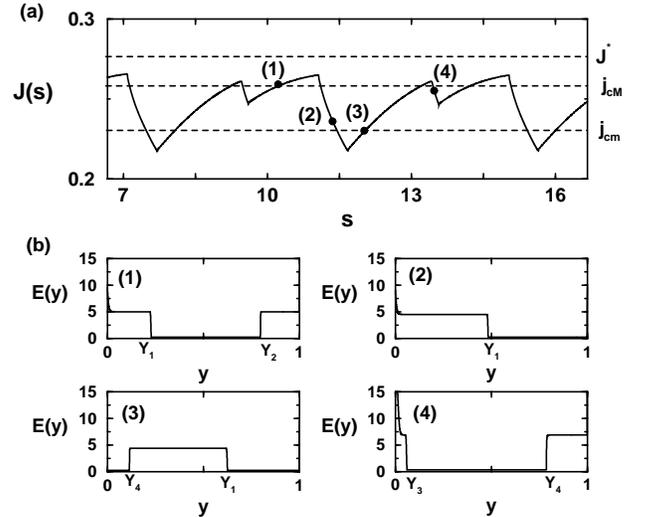}}
\caption{Gunn effect mediated by low-field dipole solitary waves. 
(a) Dimensionless 
current density $J(s)$. Parameter values are $L=1500$, $i_{0} = 0.24$,
 $\alpha_{0}=30$
and $\phi=2.3$, for which $v_{m}<j_{cm}<j_{cM}<J^{\ast}<v_{M}$.
where $J^{\ast}=0.28$ corresponds to the solution of 
$c_+(J)=c_-(J)$. The values of $j_{cM}=0.26$ and $j_{cm}=0.23$.(b) The 
corresponding electric 
field profiles $E(y,s)$ evaluated at the times marked in Part 
(a) of this figure. 
}
\label{diplow}
\end{figure}

\begin{figure}    
\centerline{
    \epsfxsize=8 cm
    \epsffile{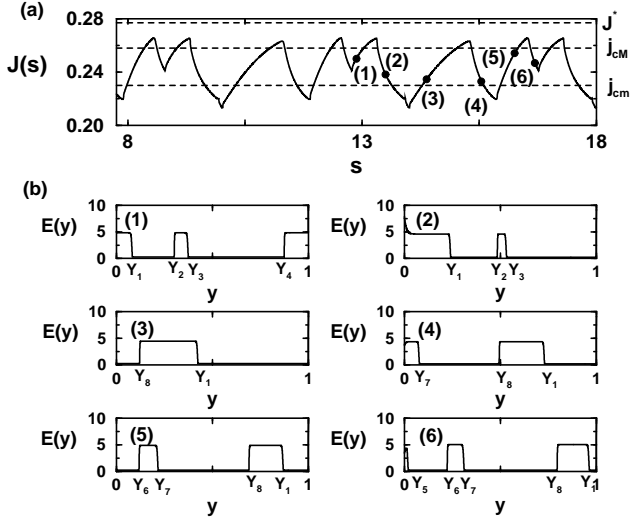}}
\caption{Multiple shedding of low-field domains. The parameter values are
$L=800$, $i_{0}=0.24$, $\alpha_{0}=30$ and $\phi=1.5$. The critical 
currents are $J^{\ast}=0.28$, $j_{cM}=0.26$ and $j_{cm}=0.23$.(a)
Dimensionless current density $J(s)$.  (b) The corresponding electric 
field profiles $E(y,s)$ evaluated at the times marked in Part 
(a) of this figure. Two pulses are 
formed during each period. The secondly shed pulse reaches and overtakes 
the first one. }
\label{multi}
\end{figure}

Low-field dipole domains appear when $v_{m}<j_{cm}<j_{cM}<J^*$ (fig.).
Depending on the value of the applied bias, single (high bias values) 
or multiple (low bias values) propagating low field domains are obtained 
[see Figs.\ \ref{diplow}, \ref{multi}].

Let us start considering the case of single
propagating low field dipole domains, Fig.\ \ref{diplow}. 
We consider an initial field configuration having a low field domain 
far from the contacts. Then there is a wavefront located
at $Y_1$, moving with speed $c_{-}$, and a wavefront located at 
$Y_2>Y_1$, moving with speed $c_{+}$, that is,
 $n_{+}=n_{-}=1$, Fig.\ \ref{diplow}b.1,
and $J(0) \in (j_{cm},J^*)$. The current increases towards $J^*$ according 
to Eq.\ (\ref{d3}), until $Y_2 =1$. Then $J$ starts satisfying (\ref{dn01}) 
(corresponding to $n_{+}=0$ and $n_{1}=1$, see Fig.\ \ref{diplow}b.2), 
and it decreases until the value $j_{cm}$ is reached. After this occurs, a 
new front (moving at speed $c_{+}$) is created, while the old wavefront 
$Y_1<1$. We have a stage described by Eq.\ (\ref{d3}) with $n_{+}=n_{-}=1$,
see Fig.\ \ref{diplow}b.3,
during which $J$ increases past $j_{cM}$ ($j_{cM}<J^*$). Then 
a new front moving with speed $c_{-}$ starts being created.
At the same time, the old front located at $Y_{1}$ reaches the end of the
sample, giving rise to a complex stage in which the non-stationary effects
can not be neglected Fig.\ \ref{diplow}b.4. After that stage, 
we recover the first situation Fig.\ \ref{diplow}b.1, and a complete period of
the oscillation has been described.  

Let us now consider an example of multiple propagating low-field
dipole domains (Fig.\ \ref{multi}). As we mentioned above, they appear for 
small bias values. The asymptotic analysis of this case is more 
complicated because there are new stages (fusion of two 
wavefronts inside the sample) whose detailed description is
outside the scope of this paper. Let us start with the configuration shown 
in Fig.\ \ref{multi}b.1, for which two low field domains 
coexist. This configuration corresponds to $n_{+}=n_{-}=2$. Following
Eq.\ (\ref{dng}), the current will evolve according to
\begin{equation}
{dJ\over ds} = 	A(J)\, [2 c_+ (J)- 2 c_- (J)]\,,
\end{equation}
thereby increasing towards $J_{2,2}=J^{*}$. Before this value can be
reached, the front located at $Y_{4}$ reaches the end of the sample. 
The resulting configuration (Fig.\ \ref{multi}b.2) has now $n_{+}=1$ and 
$n_{-}=2$, and the current decreases according to 
\begin{equation}
{dJ\over ds} = A(J)\, [c_+(J) - 2 c_-(J)]. \label{d8}
\end{equation}
If the curve $v(E)$ were 
such that the fixed point of Eq.\ (\ref{d8}) existed, the current would 
tend to such a value. For our choice of $v(E)$, this fixed point does
not exist (see Fig.\ \ref{c-pm}), and therefore the current decreases all 
the time. During this process $j_{cm}$ will be crossed. According to 
our previous considerations, a new front moving at 
speed $c_{+}$ should then be formed. However, this does not occur 
because melting of the fronts $Y_2$ and $Y_3$ starts after $Y_4$
reaches $y=1$. This melting process seems to inhibit the creation
of new wavefronts until it is completed, which happens for 
$J<j_{cm}$. Then a new front moving at speed $c_{+}$ 
is rapidly created and we have a stage, 
with $n_{+}=n_{-}=1$, Fig.\ \ref{multi}b.3, described by Eq.\ (\ref{d3}). 
The current increases until $J=j_{cM}$, at which time a front moving at 
speed $c_{-}$ appears at $y=0$, Fig.\ \ref{multi}b.4. Then the current 
decreases following (\ref{d8}) until $J=j_{cm}$. Another front moving at speed 
$c_{+}$ is then formed. We have now $n_+ = n_- = 2$, Fig.\ \ref{multi}b.5
and $J$ increases according to (\ref{d3}). When $J>j_{cM}$, a front 
moving with speed $c_-$ is created at $y=0$, so that $n_+ = 2,\, 
n_- = 3$, see Fig.\ \ref{multi}b.6. Now the current decreases towards $J_{2,3}$
according to the equation  
\begin{equation}
{dJ\over ds} = 	A(J)\, [2 c_+ (J)- 3 c_- (J)]\,.
\end{equation}
This stage lasts until $Y_1=1$, at which time we are back at 
the first stage, Fig.\ \ref{multi}b.1, having completed a full period of the 
oscillation. More complicated examples of multiple low
field domains exist and they could be described similarly. 

\subsubsection{Dynamics of charge monopoles}
When $j_{cM} > v_M$, or when $j_{cm}<v_{m}$ and $j_{sat}>v_m$, 
the Gunn effect is mediated by wavefronts 
which are charge monopoles.
The case $j_{cM} > v_M$, corresponds to moving accumulation layers,
Fig.\ \ref{monacu}, while moving depletion layers are observed for $j_{cm}<
v_{m}$ and $j_{sat}>v_m$, Fig.\ref{mondep}. H. Kroemer~\cite{Kroemer} 
discovered numerically the Gunn effect mediated by charge accumulation 
monopoles 
for boundary conditions corresponding to the control characteristics, 
while its asymptotic analysis (for $B\ll 1$) was performed many years 
later~\cite{HB}. Recently a Gunn effect mediated by charge accumulation 
monopoles has been used to describe self-sustained oscillations of the
current in GaAs/AlAs superlattices~\cite{kastrup} in the limit of 
weakly coupled, low doped, long superlattices~\cite{BKMV}. 

The complete asymptotic
analysis found in Ref.\ \cite{BKMV} can be used to describe the present
situation after a few simple changes are made. First of all the equal-areas
rule in the case of the superlattice refers to $1/v(E)$, not
$v(E)$~\cite{ICPF94}. The second important change is that of the 
injecting boundary condition in the description of the birth of a new 
monopole: instead of a rigid Neumann boundary condition $\partial 
E(0,t)/\partial x = c$ we should use (\ref{bc02}). These two changes can be 
implemented without difficulty and the details will be omitted. An important
difference with the Gunn effect mediated by dipole solitary waves is that 
the amplitude of the current oscillations is larger (its largest value is 
approximately $v_M-v_m$ for the charge accumulation monopoles and 
$j_{cM}-j_{cm}$ for the dipoles). The monopoles ``probe'' the full region 
of negative slope of $v(E)$ while the dipoles ``probe'' a smaller region.

\section{Conclusions} 
We have performed an asymptotic and numerical analysis of the Gunn 
effect in n-GaAs under general boundary conditions for 
metal-semiconductor contacts. We have shown that 
the Gunn effect is mediated by (i) moving depletion charge
monopoles, (ii) moving accumulation charge monopoles, (iii) high-field 
dipole solitary waves or (iv) low-field dipole solitary waves,
 according to whether the critical contact currents 
$j_{cM}=j_c(E_M)$, $j_{cm}=j_c(E_m)$, $j_0^{sat}$ are
(i) $j_{cm}<v_{m}$ and $j_0^{sat}>v_{m}$, (ii) $j_{cM}>v_M$, (iii) 
$v_m<j_{cm}$ and $J^*<j_{cM}<v_M$, or (iv)
$v_m<j_{cm}<j_{cM}<J^*$, in dimensionless units [$v_m$ and 
$v_M$ are the minimum and maximum values of the electron drift velocity
$v(E),\, E>0$, and $J^*$ is the solution of $c_{+}(J)=c_{-}(J)$]. Some of 
these results are well-known for boundary conditions
given by Kroemer's control characteristic. In addition we have shown
that there are new instability mechanisms consisting of multiple 
generation of (low-field) charge dipole solitary waves in the region near 
the injecting contact 
if $j_{cM}$ is close enough to $J^*$, and the dimensionless length is
large enough. In each case we have been able to describe in detail the
shape of the current oscillation,  identifying some of its main features
(maxima, minima, plateaus, \ldots) with critical currents appearing
in our model, among them the contact currents $j_{cm}$ and $j_{cM}$. This
result opens the possibility of determining contact parameters by means
of the analysis of the shape of the current oscillation. Our results 
might be of use in the analysis of self-sustained current oscillations in
weakly coupled n-doped superlattices \cite{kastrup}, once the role 
of contacts and the boundary conditions they generate are understood 
in these systems. Beside this, our results are not restricted to the 
particular model of the Gunn effect studied here, but seem to hold for a 
general class of models, supporting the idea of the ``universality''
of the Gunn effect~\cite{univ}.

\section{ACKNOWLEDGMENTS}
\label{acknowledgements} 
We thank Dr.\ M.\ Bergmann for fruitful conversations and acknowledge 
financial support from the the Spanish DGICYT through grant  PB94-0375 and
PB95-0881 and from the EC Human Capital and Mobility Programme 
contract ERBCHRXCT930413. One of
us (G.G.) acknowledge support by Generalitat de Catalunya.

\appendix

\setcounter{equation}{0}
\section{Nonstationary field at the injecting contact} \label{carac}
We shall now prove that the field $E_0(t)$ at the contact $x=0$ obeys 
the equation:
\begin{equation}
{dE_{0}\over dt} = J - j_c(E_0), \label{E_0b}	
\end{equation}
This can be seen 
from the method of characteristics applied to (\ref{eq}) with $\delta=0$ 
and the boundary condition (\ref{bc02}). The characteristic equations are
\begin{eqnarray}
{dE\over dt} = J - v(E), \label{charE}\\
	{dx\over dt} = v(E), \label{charx}
\end{eqnarray}
to be solved with the conditions
\begin{eqnarray}
E(\tau;\tau) = E_0(\tau), \label{icE}\\
x(\tau;\tau) = 0. \label{icx}
\end{eqnarray}
Clearly,
\begin{eqnarray}
{dE_{0}\over d\tau} = {\partial E\over \partial t}(\tau;\tau) 
+ {\partial E\over \partial \tau}(\tau;\tau) \nonumber\\
= J - v(E_0) + {\partial E\over \partial \tau}(\tau;\tau)\, .
 \label{ch1}
\end{eqnarray}
The last term in this equation can be obtained from the boundary condition
(\ref{bc02}) as follows. From the solution of (\ref{charx}) and (\ref{icx}),
we obtain ${\partial x\over \partial \tau}(\tau;\tau) = - v(E_0(\tau))$. 
Then the boundary condition (\ref{bc02}) yields:
\begin{eqnarray}
{\partial E\over \partial \tau}(\tau;\tau) = {\partial E\over \partial 
x}(0,\tau)\, {\partial x\over \partial \tau}(\tau;\tau)  \nonumber\\
= - v(E_0(\tau))\,\alpha_0\, (i_0 - J + {dE_{0}\over d\tau})\, . 
\label{ch2}
\end{eqnarray}
Insertion of (\ref{ch2}) into (\ref{ch1}) yields (\ref{E_0b}) (with 
$\tau = t$).

\setcounter{equation}{0}
\section{Calculation of the current, $J^{(1)}(\sigma)$,
during the shedding stage}
\label{shedding}
To obtain the equations governing the shedding stages, we need to 
keep terms of order $\epsilon$ in our calculations, as indicated 
in Ref.~\cite{BHHKV}. The outer (bulk) expansion is
\begin{eqnarray}
E = E_{i}^{(0)}(\sigma) + \epsilon E_{i}^{(1)}(\sigma) + O(\epsilon^2),
\label{e-outer}\\
J = J^{(0)}(\sigma) + \epsilon J^{(1)}(\sigma) + O(\epsilon^2),
\label{j-outer}
\end{eqnarray}
where $i=1$ if $0<y<Y_1(s_1)$ and $i=3$ if $Y_1(s_1)<y<1$. Then 
$E_i^{(0)}$ and $J^{(0)}$ solve
\begin{eqnarray}
{\partial E_{i}^{(0)}\over\partial \sigma} + v(E_i^{(0)})  = J^{(0)}(\sigma) \,,
\label{out1}\\
\phi = E_1^{(0)} + [E_3^{(0)} - E_1^{(0)}]\, [1-Y_1(s_1)] .
\label{out2}
\end{eqnarray}
We can obtain easily $E_3^{(0)}$ and $J^{(0)}$ from these equations:
\begin{eqnarray}
E_3^{(0)} = {\phi - E_{1}^{(0)} Y_{1}(s_{1}) \over 1-Y_{1}(s_{1})} \,,
\label{out3}\\
J^{(0)}(\sigma) =  v(E_1^{(0)}) Y_1(s_1) +  v(E_3^{(0)}) [1-Y_1(s_1)]\,.
\label{out4}
\end{eqnarray}
Now $E_{1}^{(0)}>$ follows from (\ref{out1}) and these two last equations.
These equations should be solved with the matching conditions that
$E_{i}^{(0)} \to E_i(j_{cM})$ ($i=1,3$) and $J^{(0)}\to j_{cM}$ as $\sigma\to 
- \infty$. These problems have the solutions $J=j_{cM}$ and $E_i^{(0)} = 
E_i(j_{cM})$. Thus we expect that the current density does not depart
substantially from $j_{cM}$ as $\epsilon\to 0$.

The $O(\epsilon)$ corrections $E_{i}^{(1)}(\sigma)$ obey the equations: 
\begin{equation}
\Theta_i E_{i}^{(1)}\equiv {\partial E_{i}^{(1)}\over\partial \sigma} 
+ v'(E_i) E_{i}^{(1)} = J^{(1)}(\sigma) \,, \label{field-1}
\end{equation}
which immediately follow from (\ref{s-eq}). To find $J^{(1)}$ we proceed as 
follows. First of all, we write the bias condition (\ref{s-bias}) 
including terms of order $\epsilon$ in the approximation of the electric 
field and the current density:
\begin{eqnarray}
\phi = E_1(j_{cM}) + [E_3(j_{cM}) - E_1(j_{cM})]\, [1-Y_1(s_1)] \nonumber\\
+\epsilon\,\left\{ E^{(1)}_1 + [E^{(1)}_3 - E^{(1)}_1]\, [1-Y_1(s_1)] 
- [E_3(j_{cM})  \right.\nonumber\\
\left.    - E_1(j_{cM})] c_+ \sigma
+ \int_0^\infty [E(x,\sigma) - E_1(j_{cM})]\, dx \right.\nonumber\\
\left. + \int_{-\infty}^0 [E(\xi) - E_1(j_{cM})]\, d\xi \right.\nonumber\\
\left. + \int_0^\infty [E(\xi) - E_3(j_{cM})]\, d\xi \right\}  +
O(\epsilon^2). \label{bias2}
\end{eqnarray}
Here $E(\xi)$ is the field inside the wavefront at $Y_1$ and $E(x,\sigma)$
is the field in the injecting layer. Since $\phi
= E_1(j_{cM}) + [E_3(j_{cM}) - E_1(j_{cM})]\, [1-Y_1(s_1)]$, we obtain 
\begin{equation}
E^{(1)}_1 + [E^{(1)}_3 - E^{(1)}_1]\, [1-Y_1(s_1)] = h(\sigma)
 \,, \label{bias-1}
\end{equation}
where $h(\sigma)$ is given by (\ref{def:h}).

We now obtain the first order differential equation (\ref{eq:J1}) for 
$J^{(1)}(\sigma)$ by applying the operator $\Theta_1 \Theta_3$ to both 
sides of (\ref{bias-1}) and then using (\ref{field-1}):
\begin{eqnarray}
\left( {\partial \over\partial\sigma} + \beta \right) [J^{(1)} - h'(\sigma) 
- \alpha h] = - \gamma h, \nonumber
\end{eqnarray}
where the coefficients $\alpha$, $\beta$ and $\gamma$ are 
\begin{eqnarray}
\alpha = v'_1 Y_1(s_1) + [1-Y_1(s_1)] v'_3 ,\nonumber\\
 \beta = v'_3 Y_1(s_1) +  [1-Y_1(s_1)] v'_1, \nonumber\\
\gamma = Y_1(s_1) [1-Y_1(s_1)] (v'_1 - v'_3)^2 .\label{alpha}
\end{eqnarray}
All functions of $J$ in these equations are calculated at $J=j_{cM}$.
Solving this equation we obtain the following function 
\begin{eqnarray}
J^{(1)}(\sigma)  = h'(\sigma) + \alpha h(\sigma) - \gamma 
\int_{-\infty}^{\sigma} e^{- \beta (\sigma - t)}\, h(t)\, dt. \label{J1}
\end{eqnarray}

The only missing function is the field at the injecting boundary layer.
This field profile is the solution of the semi-infinite problem (\ref{eq-shed}) 
to (\ref{matching}), with $J(\sigma;\epsilon) = j_{cM} + 
\epsilon J^{(1)}(\sigma)$ given by (\ref{J1}). We can write (\ref{J1}) in 
another form that suggests a more transparent interpretation:
\begin{eqnarray}
J^{(1)}(\sigma) = {\alpha\over\beta}\, \left\{ J'(s_{1})\, (\sigma - \sigma_0) 
- I'(\sigma)  \right.\nonumber\\
\left.  - \int_{0}^{\infty} \left[\alpha I'(\sigma-\sigma') 
+ v'_1 v'_3 I(\sigma-\sigma')\right]\,
e^{-\beta\sigma'}\, d\sigma' \right\} , \label{J1bis}\\
I(\sigma) = {\beta\over\alpha} \int_0^{\infty} [E(x,\sigma) - E_1]\, dx, 
\label{I}\\
\sigma_0 = -{v_{3}^{'2} Y_{1}(s_{1}) + v_{1}^{'2} [1-Y_{1}(s_{1})]
\over v'_{1} v'_{3} \beta } \nonumber\\
+ \{ [E_{3}(j_{cM}) - E_{1}(j_{cM})]\, c_{+}\}^{-1}\nonumber\\
\times \left\{ \int_{-\infty}^{0} [E(\xi) - E_{1}(j_{cM})] d\xi\right.
 \nonumber\\ 
\left. + \int_{0}^\infty [E(\xi) - E_{3}(j_{cM})] d\xi\right\}. 
\label{tau_0}
\end{eqnarray}
The terms on the right side of (\ref{J1bis}) clearly display the balance 
between the area lost by the motion of the old wavefront at $Y_1(s)$ and 
the excess area under the injecting layer. $J^{(1)}(\sigma)$ increases 
linearly with $\sigma$ until the excess area under the injecting layer
increases with $\sigma$ at least linearly.

\end{multicols}

\end{document}